\documentclass[twocolumn]{aa} 

\usepackage{graphicx}
\usepackage{txfonts}
\begin{document}

\title{On the use of C-stat in testing models for X-ray spectra}

\author{J.S. Kaastra\inst{1,2,3}}

\institute{SRON Netherlands Institute for Space Research, Sorbonnelaan 2,
           3584 CA Utrecht, the Netherlands 
           \and
           Leiden Observatory, Leiden University, PO Box 9513, 
           2300 RA Leiden, the Netherlands
           \and
           Department of Physics and Astronomy, Universiteit Utrecht, 
           P.O. Box 80000, 3508 TA Utrecht, the Netherlands
         }

\date{Received 15 July 2016; accepted 25 July 2017}

 
  \abstract
   {It has been shown that for the
analysis of X-ray spectra the 
   C-statistic, contrary to the $\chi^2$-statistic, provides unbiased
   estimates of the model parameters and their uncertainty ranges.}
   {However, it is often stated that the C-statistic 
   cannot be used to carry out statistical tests on the
   goodness of fit of the model, and therefore several investigations
   are still based on $\chi^2$-statistics.}
   {Here we show that it is straightforward to calculate the expected
   value and variance of the C-statistic so that it can be used in tests.}
   {We provide formulae and simple numerical approximations to evaluate 
   these expected values and variances. We also give examples indicating
   that tests based on only the expected value and variance of the C-statistic are
   reliable for spectra even with only $\sim$30 counts.}
   {The C-statistic can be used for statistical tests such as assessing
   the goodness of fit of a spectral model.}


   \maketitle
%

\section{Introduction}

X-ray spectra of astrophysical sources are often characterised by relatively
low numbers of counts per spectral bin. In the early days, spectral 
models were often tested using $\chi^2$-statistics. The goodness of fit
is expressed as
\begin{equation}
\chi^2 = \sum\limits_{i=1}^{n} \frac{(N_i - s_i)^2}{\sigma_i^2},
\label{eqn:chi}
\end{equation}
where the summation over $i$ is over all $n$ bins of the spectrum, $N_i$ is the
observed number of counts, $s_i$ is the expected number of counts for the tested
model, and $\sigma_i^2=s_i$ for Poissonian statistics. There are three important
remarks to make here. 

First, both the model $s_i$ and the observed spectrum $N_i$ should include the
source plus background counts to properly use Poissonian
statistics. 

Secondly, minimisation of $\chi^2$ to obtain the best-fit parameters of the
model is easier when $\sigma_i^2$ is approximated by $N_i$, which is a
reasonable approximation when $N_i$ is large and the Poissonian distribution
approaches a normal distribution, but it fails for small $N_i$, which can be
easily seen by putting $\sigma_i=0$ in (\ref{eqn:chi}). This method leads to
biased results, even for higher count rates
\citep[e.g.][]{nousek1989,mighell1999}. There are methods to compensate for
this, however all of these have some drawbacks. For example the
\citet{gehrels1986} approximation for $N_i$ in the denominator of
(\ref{eqn:chi}) causes problems when spectra are rebinned. Using
$\sigma_i^2=s_i$ \citep[e.g.][]{wheaton1995} properly requires multiple passes
through the minimisation procedure, updating $\sigma_i^2$ at each step and
sometimes leading to oscillatory behaviour.

Finally, it is often common practice in case of small $N_i$ to rebin the
spectra to at least 25 counts per bin or a similar sized number. This 
has the risk of washing out spectral details.

It has been pointed out by \citet{cash1979} that 
\begin{equation}
\tilde{C} = 2 \sum\limits_{i=1}^{n} s_i - N_i \ln (s_i)
\label{eqn:cash}
\end{equation}

is a much better statistic and can be applied to bins with a small number of
counts without any bias in the derived parameters. Also, this statistic
can be used to derive uncertainty ranges on the parameters of the model. It is
often stated that the Cash-statistic $\tilde{C}$ cannot be used to measure the
goodness of the fit using a quantity corresponding for instance to the reduced
$\chi^2_r\equiv\chi^2/(n-p),$ where $p$ is the number of free parameters. For
example, Nousek \&\ Shue (1989) noted, `The principal disadvantage of the C
statistic is that there is no value corresponding to the reduced $\chi^2$ value
with which we can measure the goodness of the fit', and Humphrey et al. (2009)
wrote, `Since the absolute value of the C-statistic cannot be directly
interpreted as a goodness-of-fit indicator observers typically prefer
instead to minimize the better-known $\chi^2$--fit statistic'. Maybe because of
this many X-ray astronomers keep using $\chi^2$-statistics, even in cases where
the approximation breaks down or leads to biased parameters.

In this paper I present calculations that can be used to evaluate the expected
value for $C$ and its root-mean-squared (rms) deviation to assess the
goodness of fit of a spectral model derived from a best fit to the observed
spectrum.

I use a modification of the original Cash-statistic that has been attributed to
Castor and is implemented in current fitting packages such as XSPEC
\citep{arnaud1996}, SHERPA \citep{freeman2001}, and SPEX \citep{kaastra1996}.
This modified C-statistic, designated here as \textsl{cstat,} is defined as
\begin{equation}
C = 2 \sum\limits_{i=1}^{n} s_i - N_i + N_i \ln (N_i/s_i)
\end{equation}
and has similar properties as the original Cash statistic, but in addition it
can be used to assign a goodness-of-fit measure to the fit.
For a spectrum with many counts per bin $C\rightarrow\chi^2$,
but where the predicted number of counts per bin is small, the expected value
for $C$ can be substantially smaller than the number of bins $n$.

\section{Expected value of $C$ and its variance\label{sect:cexp}}

\begin{figure}[!ht]
\resizebox{\hsize}{!}{\includegraphics[angle=-90]{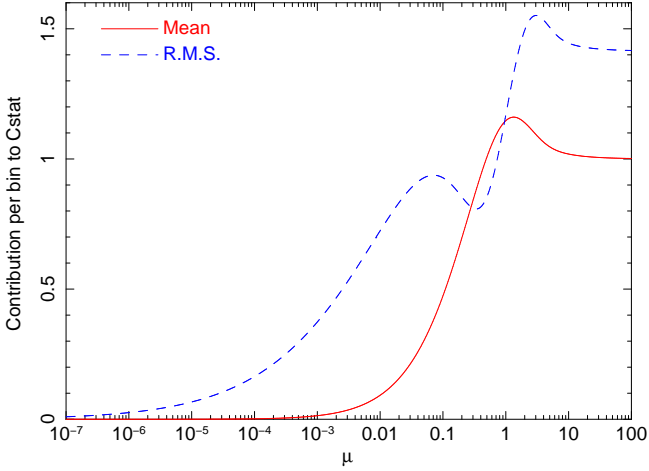}}
\caption{Expected value of the contribution per bin to $C$, and its rms
uncertainty as a function of the mean expected number of counts $\mu$.}
\label{fig:cstat}
\end{figure}

The expected contribution $C_{{\rm e},i}$ to the total $C$ from any
individual bin $i$ and its variance $C_{{\rm v},i}$ are given by
\begin{equation}
C_{{\rm e,i}} = 2 \sum_{k=0}^\infty P_k(\mu) \left[ \mu - k + k \ln (k/\mu)\right],
\label{eqn:cexp}
\end{equation}
\begin{equation}
S_{{\rm v},i} = 4 \sum_{k=0}^\infty P_k(\mu) \left[ \mu - k + k \ln (k/\mu) \right]^2,
\end{equation}
\begin{equation}
C_{{\rm v},i} = S_{{\rm v},i} - C_{{\rm e},i}^2,
\label{eqn:cvar}
\end{equation}
with $P_k(\mu)$ the Poisson distribution
\begin{equation}
P_k(\mu) = {\rm e}^{\displaystyle{-\mu}} \mu^k / k!
\end{equation}
and $\mu$ the expected number of counts in the relevant bin. We show
$C_{{\rm e},i}$ and $\sqrt{C_{{\rm v},i}}$ in Fig.~\ref{fig:cstat}.

The above equations hold for a single bin. For the full spectrum, the expected
values $C_{{\rm e},i}$ and variances $C_{{\rm v},i}$ can be simply added over
all bins $i$, yielding the expected value and variance for the full spectrum.

While the above procedure yields the exact expected mean value and variance for
$C$, in general this does not mean that $C$ has a Gaussian distribution with
that mean and variance. A Gaussian distribution occurs when each bin, or most
bins, $i$, have enough counts such that the distribution of $C_i$ becomes
asymptotically Gaussian, or for lower numbers of counts when the number of
spectral bins is large enough. In the latter case we can use the central limit
theorem, which states that when independent random variables are added, their
sum tends towards a normal distribution even if the original variables
themselves are not normally distributed.

In the above case, acceptable spectral models typically have $\Sigma C_{{\rm
e},i}(\mu_i) - f \left[\Sigma C_{{\rm v},i}(\mu_i)\right]^{0.5} < C < \Sigma
C_{{\rm e},i}(\mu_i) + f \left[ \Sigma C_{{\rm v},i}(\mu_i)\right]^{0.5}$ with
$f$ a factor of order unity corresponding to the required significance level,
for instance $f=1$ for 68\% confidence.

When these above conditions are not met, i.e. for low number of counts or low
number of spectral bins, the distribution of $C$ is not Gaussian. In that case
the higher order moments of the distribution of $C$ can be calculated
analogous to (\ref{eqn:cexp}) and (\ref{eqn:cvar}). These can be used
in principle to build the distribution of $C$. This can be a cumbersome task,
however, and alternatively, using the best-fit model, one may test it simply by
running multiple simulations of the spectrum to obtain an empirical distribution
of $C$ from which the goodness of fit can be estimated.

Fortunately, in most practical cases using the total mean and variance of $C$
with a simple Gaussian approximation is accurate enough to assess the goodness
of fit. We illustrate this with two practical examples in
Sect.~\ref{sect:example}.

We implemented the above approach in the SPEX
package\footnote{\url{www.sron.nl/spex}} \citep{kaastra1996}. To help the user
to see if a $C$-value corresponds to an acceptable fit, SPEX gives, after
spectral fitting, the expected value of $C$ and its rms spread, based on the
best-fit model. Both quantities are simply determined by adding the expected
contributions and their variances over all bins.

\section{Simple approximations for the expected value and variance of $C$\label{sect:example}}

We obtained simple approximations to the infinite series involved in
(\ref{eqn:cexp}) and (\ref{eqn:cvar}), with relative errors better than
$2.2\times 10^{-4}$ for $C_{\rm e}$ and better than $1.6\times 10^{-4}$ for
$C_{\rm v}$, as follows:

\begin{eqnarray}
0 \le \mu \le 0.5: &C_{\rm e} &= -0.25\mu^3 + 1.38\mu^2 - 2\mu\ln\mu \\
0.5< \mu \le 2:    &C_{\rm e} &= -0.00335 \mu^5 + 0.04259\mu^4 \nonumber \\
                              && - 0.27331 \mu^3+ 1.381 \mu^2 - 2 \mu\ln\mu \\
2 < \mu \le 5:     &C_{\rm e} &= 1.019275 +
0.1345\mu^{\displaystyle{0.461-0.9\ln \mu}}  \\ 
5 < \mu \le 10:    &C_{\rm e} &= 1.00624 + 0.604 / \mu^{1.68} \\
\mu >10:           &C_{\rm e} &= 1 + 0.1649 / \mu +0.226 / \mu^2
\end{eqnarray}

\begin{eqnarray}
0 \le \mu \le 0.1: &S_{\rm v} &=4 \sum_{k=0}^4 
        P_k(\mu) \left[ \mu - k + k \ln (k/\mu) \right]^2 \\
0.1 < \mu \le 0.2: &C_{\rm v} &= -262\mu^4 + 195\mu^3 -51.24\mu^2 \nonumber\\
           &&       + 4.34 \mu  + 0.77005 \\
0.2 < \mu \le 0.3: &C_{\rm v} &= 4.23 \mu^2 -2.8254 \mu + 1.12522 \\
0.3 < \mu \le 0.5: &C_{\rm v} &= -3.7\mu^3 +7.328 \mu^2 -3.6926 \mu \nonumber\\
                  && + 1.20641\\
0.5 < \mu \le 1:   &C_{\rm v} &= 1.28\mu^4 - 5.191\mu^3 + 7.666\mu^2 \nonumber\\
                         &&        - 3.5446\mu + 1.15431 \\
1 < \mu \le 2:     &C_{\rm v} &= 0.1125\mu^4 - 0.641\mu^3 + 0.859\mu^2 \nonumber\\
                         &&        + 1.0914\mu - 0.05748 \\
2 < \mu \le 3:     &C_{\rm v} &= 0.089 \mu^3 - 0.872 \mu^2 + 2.8422 \mu \nonumber\\
                    &&    - 0.67539\\
3 < \mu \le 5:     &C_{\rm v} &= 2.12336 \nonumber\\
                &&+  0.012202 \mu^{\displaystyle{5.717 - 2.6\ln\mu}} \\
5 < \mu \le 10:    &C_{\rm v} &= 2.05159 + 0.331 
                           \mu^{\displaystyle{1.343 - \ln\mu }}\\
\mu > 10: &C_{\rm v} &= 12 / \mu^3 + 0.79 / \mu^2 + 0.6747 / \mu + 2               
.\end{eqnarray}

With the help of the above equations, the goodness of fit for the model can be
easily assessed. Given the other properties of \textsl{cstat}, such as unbiased
parameter estimates, in almost all circumstances \textsl{cstat} is the preferred
statistic to be used and the use of $\chi^2$-statistics in X-ray spectral
analysis should be avoided.

\section{Two practical examples}

\begin{figure}[!ht]
\resizebox{\hsize}{!}{\includegraphics[angle=-90]{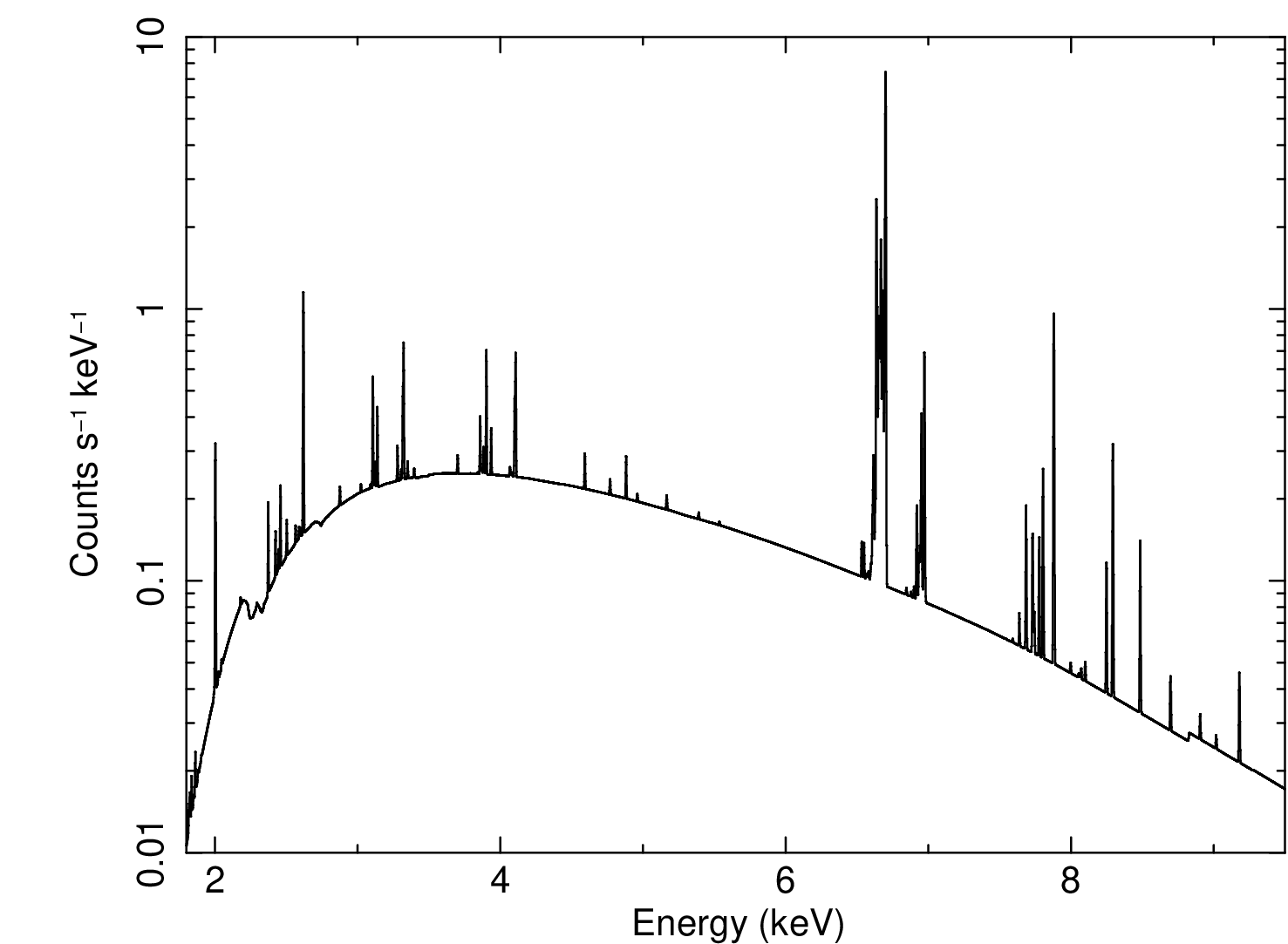}}
\resizebox{\hsize}{!}{\includegraphics[angle=-90]{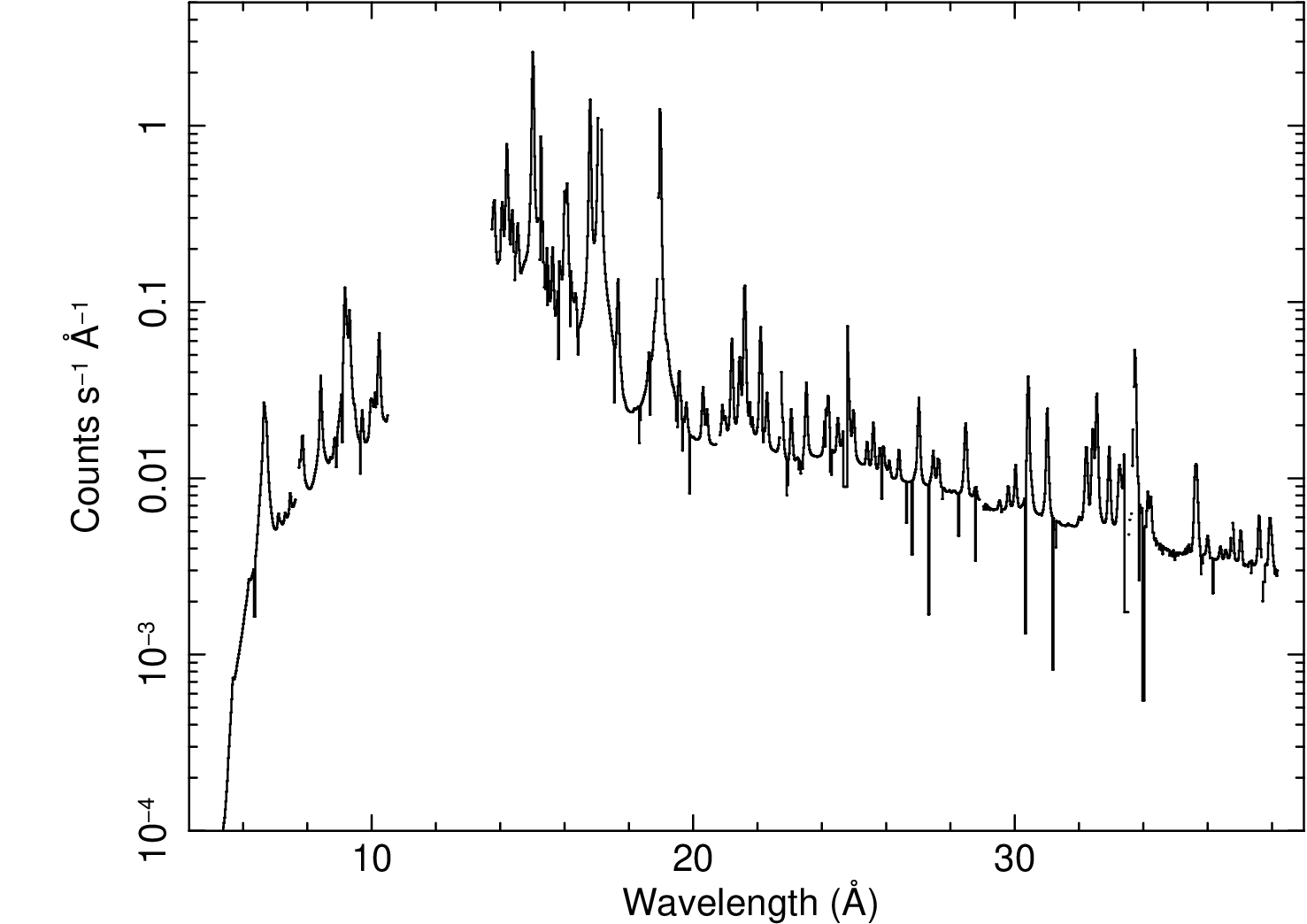}}
\caption{Simplified spectra of the Perseus cluster (top) with Hitomi and
of Capella (bottom) with RGS.}
\label{fig:spectra}
\end{figure}

We tested our method to assess the goodness of fit with two examples. In the
first example, we simulated a spectrum that has approximately the shape of the
Perseus cluster as measured with the Hitomi SXS instrument \citep{hitomi2016}.
For this demonstration purpose we simplified the model by adopting an isothermal
spectrum in collisional ionisation equilibrium with a temperature of 4~keV,
proto-solar abundances, and an emission measure matching the flux of Perseus as
measured by Hitomi. In our spectral fits, we used the temperature and emission
measure of the source as free parameters. The spectrum has 5804 bins and is
shown in Fig.~\ref{fig:spectra}, without noise, but with instrumental features
included.

\begin{figure}[!ht]
\resizebox{\hsize}{!}{\includegraphics[angle=-90]{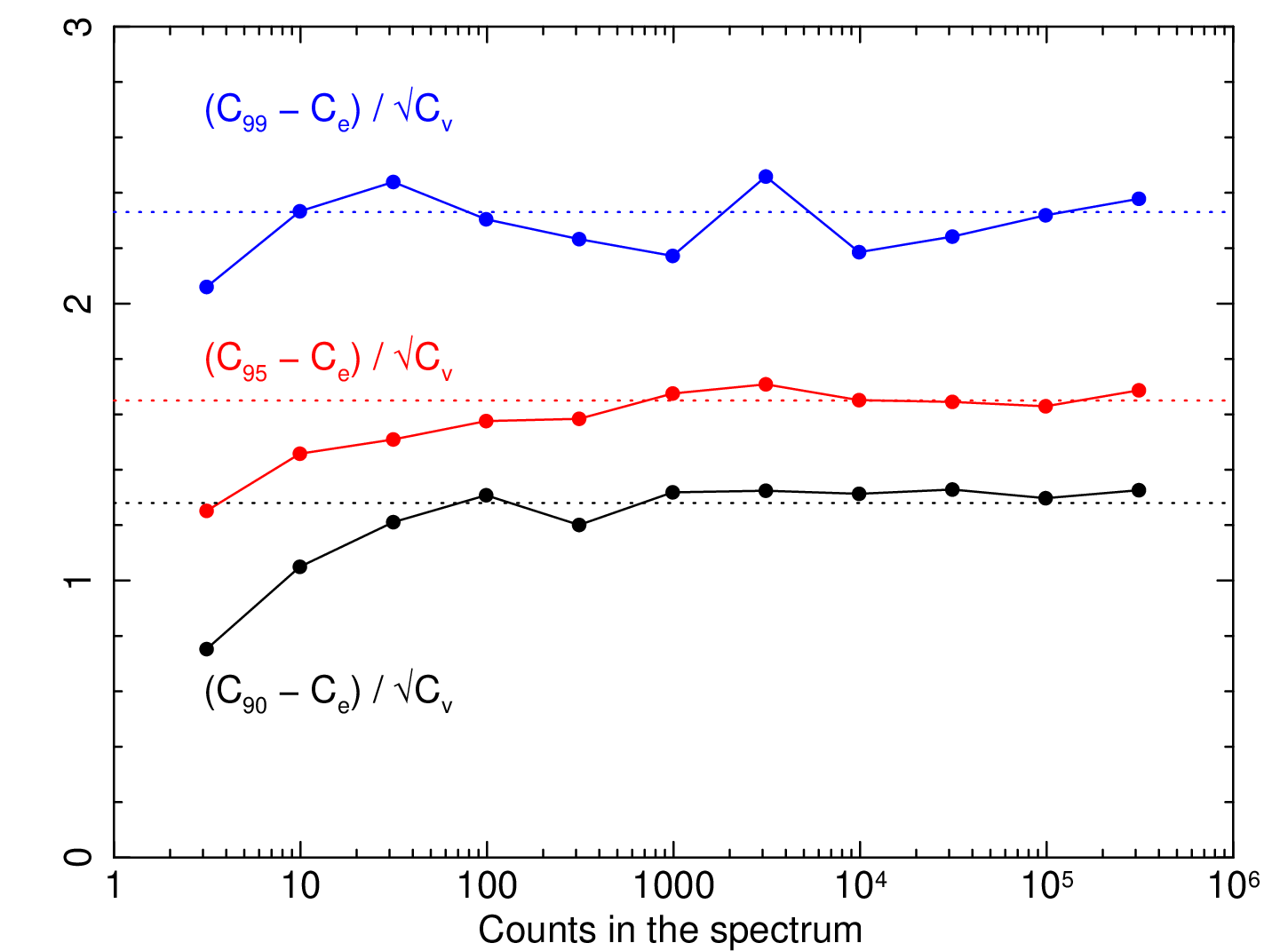}}
\resizebox{\hsize}{!}{\includegraphics[angle=-90]{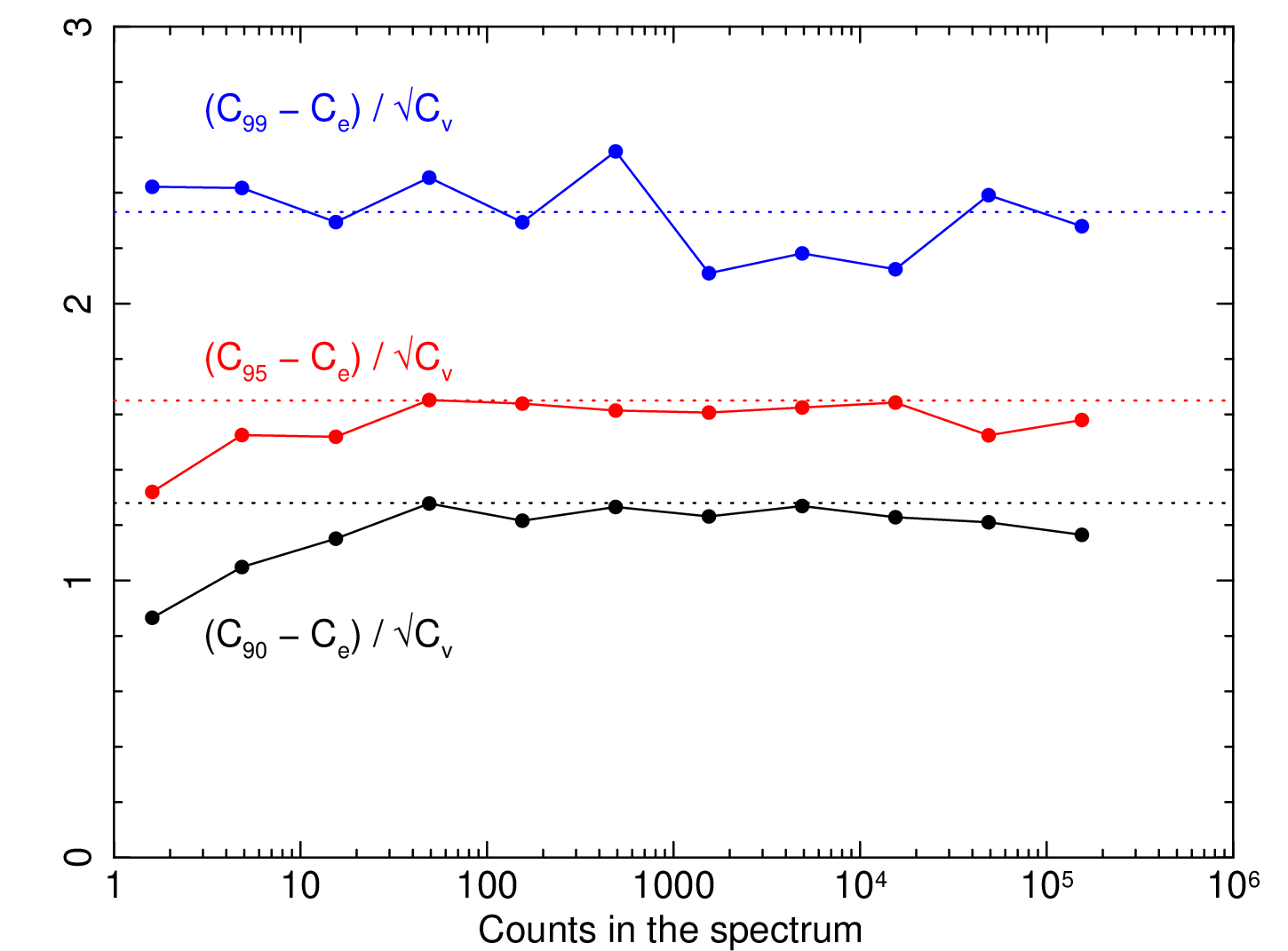}}
\caption{Percentile points 90\%, 95\%, and 99\% for the distribution of $C$ for
simulated Perseus (top) and Capella (bottom) spectra. The expected value $C_{\rm
e}$ was subtracted and the difference is scaled with $\sqrt{C_{\rm v}}$.
The percentile points are shown as a function of the average number of counts in
the spectra. The dotted lines show the expected percentile values if the
distribution of $C$ had been normal. These values are reached
asymptotically for large numbers of counts in the spectra.}
\label{fig:percentiles}
\end{figure}

We scaled this spectrum in flux by factors of $10^{-k/2}$ with $k$ ranging from
0--10, i.e. higher values of $k$ corresponding to lower fluxes. For each flux
value, we simulated 1000 spectra, performed a best fit, and determined $C$. We
then produced a histogram of $C$-values and calculated the 90\%, 95\%, and 99\%
percentile points $C_{90}$, $C_{95}$, and $C_{99}$ of this distribution. In an
actual analysis of an observed spectrum, a model would be rejected at the 90\%
confidence level if $C>C_{90}$, and this is similar for the other confidence
levels. We compare these percentile points, scaled with the expected mean
$C_{\rm e}$ and variance $C_{\rm v}$, as delivered by SPEX and described in
Sect.~\ref{sect:cexp}, in Fig.~\ref{fig:percentiles}.

Our second example is a simulation of the spectrum of Capella with the
Reflection Grating Spectrometer (RGS) of XMM-Newton. This spectrum is
approximated by a single isothermal component with temperature 0.5~keV and
appropriate flux for Capella. All further steps are the same as for the Perseus
example. The main difference between both spectra is that while Perseus is
dominated by continuum emission, owing to its high temperature, Capella is
dominated by line emission, owing to its low temperature. The Capella spectrum
has 1433 bins and is also shown in Fig.~\ref{fig:spectra}.

From Fig.~\ref{fig:percentiles} we see that for more than about 30 counts in the
spectrum the percentile points $C_{90}$ and $C_{95}$ (dots connected by solid
lines) are close to the values calculated from the mean value and variance of
$C$ using a Gaussian approximation (dotted lines). This holds for both examples.
For $C_{95}$, the Gaussian approximation even works for spectra with 10 counts.

But even going down to about 5 counts does not result in dramatic differences.
In general one needs to be very cautious with spectra that have so few counts.
For instance, 30 counts correspond to an average number of counts per bin of
0.005 and 0.02 only for the scaled Perseus and Capella spectra, respectively.

\section{Conclusion}

The C-statistic can be used to estimate the goodness of fit of a model in the
vast majority of all cases. When the spectrum has more than 10--30 counts, the
distribution of $C$ for the tested model is close enough to a Gaussian
distribution for the highest confidence levels above 95\%. This paper describes
an algorithm that calculates the expected mean and variance of $C$ that can be
used to assess these confidence levels using a simple normal distribution.

\begin{acknowledgements}

SRON is supported financially by NWO, the Netherlands Organization for
Scientific Research. I thank the referee for useful comments.

\end{acknowledgements}

\bibliographystyle{aa}
\bibliography{paper}

\end{document}